\documentclass[aps,prl,twocolumn,superscriptaddress,preprintnumbers,showpacs]{revtex4}
\usepackage{graphics}
\def\eq#1\en{\begin{equation} #1 \end{equation}}

\def\eqa#1\ena{\begin{eqnarray} #1 \end{eqnarray}}

\usepackage{graphicx,color}

\begin{document}

\title{Rare $\Omega^{-} \to \Xi^{0}(1530) \pi^-$ decay in the Skyrme model}
\author{G. Duplan\v{c}i\'{c}}
\affiliation{Theoretical Physics Division, Rudjer Bo\v skovi\' c Institute, 
Zagreb, Croatia}
\author{H. Pa\v{s}agi\'{c}}
\affiliation{Faculty of Transport and Traffic Engineering, University of Zagreb, 
P.O. Box 195, 10000 Zagreb, Croatia} 
\author{J.Trampeti\'{c}}
\affiliation{Theoretical Physics Division, Rudjer Bo\v skovi\' c Institute, 
Zagreb, Croatia}
\affiliation{Theoretische Physik, Universit\"{a}t M\"{u}nchen, Theresienstr. 37, 80333 M\"{u}nchen, Germany}

\date{\today}

\begin{abstract}
Rare nonleptonic $\Omega^{-} \to \Xi^{0}(1530) \pi^-$ decay branching ratio is estimated by means
of the QCD enhanced effective weak Hamiltonian supplemented by the SU(3)
Skyrme model used to estimate the nonperturbative matrix elements.  
Using mean values for experimental input parameters and the Skyrme charge 
$e=4.75$ we obtain the rate which is in a good agreement with data.
\end{abstract}

\pacs{PACS number(s): 12.39.Dc, 12.39.Fe, 13.30.Eg}

\maketitle

It is well known that the nonleptonic weak decays of baryons can be
reasonably well described in the framework of the Standard Model \cite
{ga,do2}. 
Recently both s-, p-wave nonleptonic hyperon and p-wave 
$\Omega^{-}$ decay amplitudes were quite
successfully reproduced by the SU(3) extended Skyrme model with the QCD
enhanced effective weak Hamiltonian \cite{du,du1}. The decay amplitudes were
described through the current-algebra commutator, the ground-state baryon
pole terms and factorizable contributions. The nonperturbative quantities,
i.e. the baryon six dimensional operator matrix elements, were determined 
using the SU(3) extended Skyrme model. For the s-wave hyperon decay
amplitudes correct relative signs and absolute magnitudes were obtained. For 
the p-waves all relative signs were correct, with their relative
magnitudes roughly following the experimental data. One is thus faced with an obvious
question: could an analogous approach work equally well for the rare 
nonleptonic $\Omega ^{-}\to \Xi^{0}_{3/2^+}(1530) \pi^-$ \cite{foot1} weak decay? 
Such a question should be considered in connection
with the measurements of the $\Omega ^{-}$ lifetime and branching ratios \cite{rpp}: 
\begin{equation}
\tau (\Omega ^{-})_{\rm exp}=(82.1\pm 1.1)\;\; {\rm ps}
\label{1}
\end{equation}
\begin{equation}
\Gamma(\Omega ^{-}\to \Xi^{*0} \pi^-)/\Gamma_{tot} = 
\left(6.4 {+5.1 \atop -2.0}\right) \times 10^{-4} .
\label{2}
\end{equation}
In this report our goal is to test whether the effective weak Hamiltonian and the SU(3)
Skyrme model are able to predict the rare nonleptonic 
$\Omega ^{-}\to \Xi^{*0} \pi^-$ decay following the method 
in \cite{du,du1}. To this end we shall employ the Standard Model effective
Hamiltonian and {\it the minimal number of couplings concept} of the Skyrme
model to estimate the nonperturbative matrix elements of the 4-quark
operators \cite{du,du1,tra} for decuplet-decuplet
transitions. Throughout
this report we use the {\it arctan ansatz} for the Skyrme profile function $%
F(r)$ \cite{dia}, which allows to calculate the pertinent overlap
integrals analytically with accuracy of the order of $\stackrel{<}{\sim}$1 \% with respect
to the exact numerical results.

The starting point in such an analysis is the effective weak
Hamiltonian in the form of the current $\otimes $ current interaction,
enhanced by QCD,  
\begin{equation}
H_{w}^{eff}(\Delta S=1)=\sqrt{2}G_{F}V_{ud}^{\ast
}V_{us}^{}\sum c_{i}O_{i},  
\label{3}
\end{equation}
where $G_{F}$ is the Fermi constant and $V_{ud}^{\ast }V_{us}^{}$ are the
Cabbibo-Kobayashi-Maskawa matrix elements. The $O_{i}$ are the 4-quark operators
and the $c_{i}$ factors are the QCD-short distance
Wilson coefficients \cite{do2,bur}: 
$c_{1}=-1.90-0.61\zeta,\;\;c_{2}=0.14+0.020\zeta,
\;\;c_{3}=c_{4}/5,\;\;c_{4}=0.49+0.005\zeta $, with $\zeta =V_{td}^{\ast
}V_{ts}^{}/V_{ud}^{\ast }V_{us}^{}$. 
For the purpose of this work we neglect the so called Penguin
operators since their contributions are proven to be small \cite{bur}.  
Without QCD corrections, the
Wilson coefficients would have the following values: $c_{1}=-1,\;c_{2}=1/5,%
\;c_{3}=2/15,\;c_{4}=2/3\,$. In this paper we consider both
possibilities and compare the resulting rates.

The techniques used to describe nonleptonic $\Omega ^{-}$ decays (in this work we have only
the $3/2^{+}\rightarrow 3/2^{+}+0^{-}$ reaction) are
known as a modified current-algebra (CA) approach. The general form
of the decay amplitude reads:
\begin{eqnarray}
&&\langle \pi (q)B^{\prime }(p^{\prime })|H_{w}^{eff}|B(p)\rangle\! = 
\label{4}\\
&&={\cal {\overline W}}_{\mu }%
(p^{\prime })[({\cal A}+\gamma _{5}{\cal B})g^{\mu \nu}+
({\cal C}+\gamma _{5}{\cal D})q^{\mu} q^{\nu}]{\cal W}_{\nu }(p). 
\nonumber 
\end{eqnarray}
The ${\cal W}(p)$ denotes the Rarita-Schwinger spinor. The parity-violating amplitudes $%
{\cal A}$ correspond to the s-wave and parity-conserving amplitudes ${\cal B}$
correspond to the p-wave $\Omega ^{-}$ decays, respectively.
Since the decay of $\Xi^{*0} \pi^-$ is strongly limited by phase space
(momentum transfer $\simeq$ 1 MeV at the peak value of the $\Xi^*$ mass),
we will neglect the amplitudes $\cal C$ and $\cal D$ (d- and f- waves).
The decay probability 
$\Gamma (\Omega^-_{3/2^{+}}(p)\rightarrow \Xi^{*0}_{3/2^{+}}(p')+\pi^-_{0^{-}}(q))$ is: 
\begin{eqnarray}
\Gamma &=&\frac{|{\bf q}|m_f}{18\pi m_{\Omega }}\left\{ 
\left(\frac{E'}{m_f}+1\right)\left[\frac{E'^2}{m^2_f}+\frac{E^{\prime}}{m_f}+\frac{5}{2}\right]|{\cal A}|^{2}
\right. \nonumber \\
&+&\left. \left(\frac{E'}{m_f}-1\right)\left[\frac{E'^2}{m^2_f}-\frac{E'}{m_f}+\frac{5}{2}\right]
|{\cal B}|^{2}\right\} ,  
\label{5}
\\
|{\bf q}|^{2} &=&\left[ (m_{\Omega }^{2}-m_{f}^{2}+m_{\phi
}^{2})/2m_{\Omega }\right] ^{2}-m_{\phi }^{2},  
\label{6} \\
E^{\prime } &=&(m_{\Omega }^{2}+m_{f}^{2}-m_{\phi }^{2})/2m_{\Omega }.
\label{7}
\end{eqnarray}
Here $\Omega^-$ is at rest, $m_{f}$ denotes the final baryon mass and $m_{\phi }$ is the mass of
the emitted meson. 

We calculate the decay amplitudes by
using the so-called tree-diagram approximation at the particle
level, i.e. factorizable and pole diagrams plus the commutator term.
The amplitudes (\ref{4}) receive contributions from the commutator term
and the pole diagram, respectively: 
\begin{eqnarray}
A_{Comm} &=&\frac{-1}{{\sqrt 2} f_{\pi}} \; a_{\Xi^{*}\Omega},
\label{8}
\\
B_{{\cal P}} &=&\frac{-g_{\Xi ^{*0}\Xi ^{\ast -}\pi^{-}}}
{m_{\Omega}-m_{\Xi^{*}}} \; a_{\Xi^{*}\Omega}. 
\label{9}
\end{eqnarray}
However, another state rather strongly coupled to $\Xi^* \pi$ and with a mass close to the
$\Omega^-$ exists, the $\Xi(1820, J^P=3/2^-)$ resonance \cite{foot2}, whose mass fits nicely to the
Gell-Man-Okubo formula for an octet of $3/2^-$ baryons. Therefore, there is a pole term
in the s-wave amplitudes, $A_{\cal P}$:
\begin{eqnarray}
A_{{\cal P}} =\frac{g_{\Xi^{*}\Xi^{**}\pi}}{m_{\Omega}-m_{\Xi ^{**}}} \;
 b_{\Xi ^{**}\Omega}. 
\label{10}
\end{eqnarray}
The commutator and the baryon-pole amplitudes contain weak matrix elements
defined as 
\begin{eqnarray}
{a_{\Xi ^{*}\Omega} \choose b_{\Xi ^{**}\Omega}}
={\sqrt{2}G_F V^*_{ud}V_{us}^{}\;}   
{\langle \Xi ^{*}|c_{i}O_{i}^{PC}|\Omega \rangle 
\choose
\langle \Xi ^{**}|c_{i}O_{i}^{PV}|\Omega\rangle}, 
\label{11}
\end{eqnarray}
where the important parts are the Wilson coefficients and the 4-quark
operator matrix elements. 

The factorizable contributions to s- and p-waves, are calculated by
inserting the vacuum states; it is therefore a factorized product of two
current matrix elements, where the decuplet-decuplet matrix element of the
vector and the axial-vector currents reads: 
\begin{eqnarray}
\langle \Xi^*(p^{\prime })|{V \choose A}^{\mu }|\Omega ^{-}(p)\rangle =g_{V \choose A}^{\Xi^* \Omega }
{\cal {\overline W}}_{\nu }(p){\gamma}^{\mu}{1 \choose \gamma_5}{\cal W}^{\nu }(p'). 
\label{12}
\end{eqnarray}
Summing over all factorizable contributions gives the following expressions
for the amplitudes: 
\begin{eqnarray}
{A \choose B}_{{\cal S}}&=&\frac{G_F}{\sqrt{3}}V^*_{ud}V_{us}^{}
(m_{\Omega}\mp m_{\Xi^*})f_{\pi }
\label{13}\\
&\times& g_{V \choose A}^{\Xi^{*}\Omega}\left[ c_{1}-2(c_{2}+c_{3}+c_{4})\right], 
\nonumber
\end{eqnarray}
where the $g_{A(V)}^{\Xi^* \Omega }$ represents the form-factor of the spatial
component of the axial-vector(vector) current. 

The total theoretical amplitudes are: 
\begin{eqnarray}
{\cal A}_{\rm th}(m_{\pi}^2 )&=&A_{Comm}(0)+A_{{\cal P}}(m_{\pi}^2 )+A_{{\cal S}}(m_{\pi}^2 ),
 \label{14} \\
{\cal B}_{\rm th}(m_{\pi}^2 )&=&B_{{\cal P}}(m_{\pi}^2 )+B_{{\cal S}}(m_{\pi}^2 ), 
\label{15}
\end{eqnarray}
were the relative signs between commutator, pole and factorizable contributions
are determined via SU(3) and the generalized Goldberger-Treiman relation.

In order to estimate the 4-quark operator matrix elements entering (\ref{11}), we take
the Skyrme model where baryons emerge as soliton configurations of the field $U$ 
of pseudo-scalar mesons \cite{sky,adk,gua,yab,NMPJW,wei}. 
The SU(3) extended Skyrme model action is 
${\cal L}={\cal L}_{\sigma }+{\cal L}_{Sk}+{\cal L}_{SB}+{\cal L}_{WZ}$,
where ${\cal L}_{\sigma }$, ${\cal L}_{Sk}$, ${\cal L}_{SB}$, and 
${\cal L}_{WZ}$ denote the $\sigma $-model, Skyrme, symmetry breaking (SB), and
Wess-Zumino (WZ) terms, respectively \cite{wei,sco,DPT}. 
Extension of the model to the
strange sector \cite{gua,NMPJW,wei} is done by an isospin embedding of
the static {\em hedgehog} ansatz into an SU(3) matrix, which is a subject of
a time dependent rotation ${\rm U}({\mathbf r},t)=A(t){{\cal U}({\mathbf r})}A^{\dagger}(t)$ 
by a collective coordinate matrix $A(t)\in SU(3)$. The
generalized velocities are defined by $A^{\dagger }(t)\dot{A}(t)={\frac{i}{2}}\sum_{\alpha
=1}^{8}\lambda _{\alpha }\dot{a}^{\alpha }$ and the profile function
is interpreted as a chiral angle that parameterizes the soliton. The
collective coordinates $a^{\alpha }$ are canonically quantized to
generate the states that possess the quantum numbers of the physical strange
baryons. In order to account for a non-zero strange quark mass the
appropriate symmetry breaking terms should be included. 

In this work we will use the SU(3) extended action $\cal L$
with the following set of parameters \cite{DPT}, ${\hat{x}}=36.97$, 
$\beta ^{\prime }=-28.6\,{\rm MeV}^2$, $\delta ^{\prime }=4.12\times 10^7\,{\rm MeV}^4$, 
determined from the masses and decay constants of the pseudo-scalar mesons. 
The $\hat{x}$ term is responsible for the baryon mass
splittings and the admixture of higher representations in the baryon wave
functions. However, we use the SU(3)
symmetric baryon wave functions in the spirit of the perturbative approach
to SB. Indeed, we have shown in \cite{du,du1}, that the WZ and SB contributions through
the weak operator matrix elements are small and introduce
few \% uncertainty of the decay amplitudes dominated by
the Skyrme term which scales like  $1/e$. 

Since the coupling $f_{\pi }$ is equal to its experimental value, 
for the evaluation of rare nonleptonic $\Omega ^{-}$ decay rate
the only remaining free parameter, as in \cite{du,du1}, is the Skyrme
charge $e$. It has been shown in Figure 12 of \cite{DPT} that  
for $4.0\leq e\leq 5.0$ the mass spectrums of ${\bf 8},{\bf 10}$ and 
$\overline{\bf 10}$-plets are reasonably well described.
This is the reason we are using that particular range of $e$ further on. 

The 4-quark operator matrix element contribution to the commutator (\ref{8})
and to the baryon pole term (\ref{9}), evaluated as a function of 
the Skyrme charge $e$ \cite{du1}, in units of [$10^{-8}$ GeV], is 
\begin{eqnarray}
a_{\Xi^{*-}\Omega^{-}}=
\left\{{\begin{array}{cc|c}
{\rm QCD \;off} & {\rm QCD \;on}  & e\\
 \hline
 -1.86 & -3.52 & 4.00 \\
 -1.31 & -2.75 & 4.75\\
 -1.22 & -2.55 & 5.00  
\end{array}}\right.
\label{18}
\end{eqnarray}

The size of the strong coupling from pole term (\ref{9}),
determined via generalized Goldberger-Treiman relation,
\begin{equation}
g_{{\Xi^{*-}}{\Xi^{0*}}\pi^-}=
\frac{m_{{\Xi^{*-}}}+m_{{\Xi^{*0}}}}{2f_{\pi}} g_{A}^{{\Xi^{*-}}{\Xi^{0*}}}
\cong 9.11\,g_{A}^{\rm pn},
\label{19}
\end{equation}
is close to the estimate given in \cite{lu}.

The strong coupling $g_{\Xi^{**}\Xi^*\pi}$, which enters 
in the calculation of $A_{\cal P}$, we
extract from experimental value for partial width $\Gamma(\Xi^{**}\rightarrow \Xi^*\pi)$. 
Using strong effective Lagrangian for spin $3/2^-$ baryon $\Xi^{**}$  
\begin{eqnarray}
{\cal L}_{\rm int}(\Xi^{**}\rightarrow \Xi^*\pi)= 
g_{\Xi^{**}\Xi^*\pi}\,{\overline{\cal W}_{\mu}}(p'){\tilde{\cal W}}^{\mu}(p)\phi_{\pi}(q),
\label{21}
\end{eqnarray}
where ${\tilde{\cal W}}^{\mu}$ denotes the $3/2^-$ Rarita-Schwinger spinor, 
we obtain for $\Xi^{**}$ at rest,
\begin{eqnarray}
\Gamma(\Xi^{**}\rightarrow \Xi^*\pi) &=&
\frac{g^2_{\Xi^{**}\Xi^*\pi}|{\bf q}|}{36\pi m^2_{\Xi^*} m_{\Xi^{**}}}
(E'+m_{\Xi^*})
\label{22}\\
&\times&\left[2E'^2+2E'm_{\Xi^*}+5m^2_{\Xi^*}\right].
\nonumber
\end{eqnarray}
The $E'$ and $|{\bf q}|$ are given in (\ref{6}) and (\ref{7}).
From experiment \cite{rpp} we have $\Gamma_{\rm tot}(\Xi^{**}) = (24\pm 6) \;\rm MeV$ and
\begin{eqnarray}
\Gamma(\Xi^{**}\rightarrow \Xi^*\pi)/\Gamma_{\rm tot}(\Xi^{**}) = (0.30\pm 0.15).
\label{23}
\end{eqnarray}

The generalized Goldberger-Treiman relation applied to (\ref{10}) gives the product 
$b_{\Xi ^{**}\Omega}g^A_{\Xi ^{**}\Xi ^{*}}$ which does not depend on the phase of 
the exchanged $\Xi ^{**}$ in the pole diagram. It has been found in \cite{pak} that 
the sign of an axial-vector matrix element, between particles of the opposite parity,
is negative. Than from (\ref{22}) and (\ref{23}) we find 
$g_{\Xi ^{**}\Xi ^{*}\pi} = -\,(0.48\pm 0.14)$,
a small negative value with large uncertainty. 

The weak matrix elements of the same spin but opposite parity baryons  
are found to be about the same as the ones including ground states only \cite{pak},
i.e. $b_{\Xi ^{**}\Omega} \cong \;a_{\Xi ^{\ast}\Omega}$. 
This together with (\ref{8}) and (\ref{10}) gives
\begin{eqnarray}
A_{\cal P} = -(0.44\pm 0.12)\; A_{Comm},
\label{25}
\end{eqnarray}
up to the sign, similar to the conclusion in \cite{lu}.
Since the resonance $\Xi^{**}$ is known with quite a large error, the same error is 
also present in (\ref{25}).

We proceed with the computation of the vector and axial-vector current form-factors 
$g_{V(A)}^{\Xi^*\Omega }$, in terms of 
$g_{V(A)}^{\rm pn}$, via SU(3) Clebsh-Gordan coefficients.
Using from experiment, $g^{\rm pn}_A/g^{\rm pn}_V=1.26$, we find an agreement with 
the SU(6) result \cite{lu}, as we should: 
\begin{equation}
g_{V \choose A}^{\Xi ^{*-}\Omega ^{-}}=g_{V \choose A}^{\Xi ^{*0}\Omega ^{-}}
\cong {1.1 \choose 0.7}.
\label{20}
\end{equation}
Factorizable amplitudes $A_{\cal S}$ and $B_{\cal S}$ are very slowly changing 
functions, via the $g_A^{\rm pn}/g^{\rm pn}_V$, with respect to the Skyrme charge $e$,
(for details see Fig's 2-4 in \cite{DPT}). For QCD off/on we found from (\ref{13}) and (\ref{20}): 
\begin{eqnarray}
A^{\rm off}_{\cal S}&=&-5.65\times 10^{-8},\;
A^{\rm on}_{\cal S}=-6.32\times 10^{-8},
\nonumber
\\
B^{\rm off}_{\cal S}&=&-90.1\times 10^{-8},\;
B^{\rm on}_{\cal S}=-100.8\times 10^{-8}.
\label{26a}
\end{eqnarray}
The mean values of theoretical amplitudes $A_{Comm}$, $A_{\cal P}$ and $B_{\cal P}$ for QCD off/on, 
as a functions of the Skyrme charge $e$, are presented in Table \ref{t:tab1}. 
\renewcommand{\arraystretch}{1.4}
\begin{table}
\caption{The amplitudes $A_{Comm}$, $A_{\cal P}$, $B_{\cal P}$ in units $[10^{-8}]$ 
contributing to the rare nonleptonic two-body $\Omega^-$ decay rate.}
\begin{center}
\begin{tabular}{|c|ccc|ccc|}
\hline
$ $ & $$ & ${\rm QCD \;off}$  & $$ & $$ & ${\rm QCD \;on}$ & $$ \\
\hline\hline
$e$ 
& $A_{Comm}$ & $A_{\cal P}$  & $B_{\cal P}$  
& $A_{Comm}$ & $A_{\cal P}$  & $B_{\cal P}$ \\
\hline 
$4.00$ 
& $ 14.14$ & $-6.22 $  & $ 155.6 $ 
& $ 26.76 $ & $ -11.77 $ & $ 294.5 $\\


$4.75$ 
& $ 9.96$ & $-4.38 $  & $ 109.6 $ 
& $ 20.91 $ & $ -9.20 $ & $ 230.1 $\\

$5.00$ 
& $ 9.28$ & $-4.08 $  & $ 102.1 $ 
& $ 19.39 $ & $ -8.53 $ & $ 213.4 $\\
\hline 
\end{tabular}
\label{t:tab1}
\end{center}
\end{table}
\renewcommand{\arraystretch}{1}

Rare nonleptonic two-body $\Omega^-$ decay partial width formulae 
\begin{eqnarray}
&&\Gamma_{\rm th}(\Omega ^{-}\to \Xi^{*0} \pi^-)
=\left\{9.000617|A_{\cal S}+A_{Comm}+A_{\cal P}|^2
\right.
\nonumber\\
&&+\left.0.000148|B_{\cal S}+B_{\cal P}|^2\right\}2.697\times 10^{-4} \;\rm GeV,
\label{26}
\end{eqnarray}
and eqs. (\ref{8}) to (\ref{26a}), in units of [$10^{-18}$ GeV], gives
\begin{eqnarray}
\Gamma_{\rm th}(\Omega ^{-}\to \Xi^{*0} \pi^-)=
\left\{{\begin{array}{cc|c}
{\rm QCD \;off} & {\rm QCD \;on} & e\\
 \hline
 1.19^{+2.65}_{-1.09} & 17.81^{+16.57}_{-10.44} & 4.00\\
 0.01^{+0.38}_{-0.01} & 6.84^{+8.36}_{-4.77} & 4.75\\
 0.06^{+0.54}_{-0.06} & 4.83^{+6.65}_{-3.60} & 5.00  
\end{array}}\right.
\label{27}
\end{eqnarray}
which, compared with the experiment \cite{rpp}, 
\begin{equation}
\Gamma_{\rm exp}(\Omega ^{-}\to \Xi^{*0} \pi^-)
= \left(5.1{+4.1\atop -1.6}\right)\times 10^{-18} \; \rm GeV,
\label{28}
\end{equation}
shows the following:

(a) In this dynamical scheme framework,   
contrary to the nonleptonic hyperon and $\Omega^-$ decays \cite{du,du1}, 
the factorizable contributions turn out to be very important for the rare 
$\Omega^{-}\to \Xi^{*0} \pi^-$ nonleptonic decay. 
The opposite signs between the commutator and pole term (\ref{25})
and the factorizable contributions (\ref{26a}) becomes an essential feature, 
of our dynamical scheme, leading
to the internal cancellation within the $\cal A$ and $\cal B$ amplitudes and bringing theoretical
estimate (\ref{27}) closer to the experiment.

(b) From theoretical rate (\ref{27}) it is clear that 
in this dynamical scheme the QCD enhancement is crucial. 
  
(c) Important feature of the particular rare  
decay mode $\Omega^- \to \Xi^{*0} \pi^-$ 
lies in the fact that it is strongly limited by phase space and that 
the decay transition occurs at threshold.
Thus, $\pi^-$ in the final state is almost at rest.

(d) First consequence of (c) is the 
emitted pion is really soft and the so called soft pion limit theorem is very well satisfied.
This is the reason for the commutator term dominance in the s-wave amplitude ${\cal A}$.
Second consequence of (c) is an overall phase space  
enhancement, by factor of $6.1\times 10^4$, for the s-waves.
Owing to this, altho the p-wave amplitudes are more than one order of magnitude larger
than the corresponding s-waves, the total contribution of ${\cal B}$ amplitudes
to the rate (\ref{27}) is very small. 

(e) Inspection of the result (\ref{27}) shows the importance of the $e$ dependence and
that $e=4.75$, with QCD corrections switched on, 
gives the central fit for the rare $\Omega^{-}\to \Xi^{*0} \pi^-$ nonleptonic decay.
Applying  the same $e$ and QCD switched on to the $\Omega^{-}$ nonleptonic 
decays from \cite{du1}, we first find $a_{\Lambda \Xi^0}=4.85\times 10^{-8}$ GeV. 
This together with the matrix element $a_{\Xi^{*-}\Omega^{-}}=-2.75\times 10^{-8}$ GeV
and factorizable contributions from Table 1 in \cite{du1} produces the  
$\Omega^{-}$ nonleptonic decay amplitudes: ${\cal B}_{th}(\Omega_K)= 6.02$,
${\cal B}_{th}(\Omega^-_-)= 1.48$ and ${\cal B}_{th}(\Omega^-_0)= 1.29$, in units of
[$10^{-6} \; \rm GeV^{-1}$], which are closer to 
the conclusions in \cite{EBH} and in better agreement 
with experiment than the ones obtained in \cite{du1}. 
The same should hold for the p-wave nonleptonic hyperon decays \cite{du}. 

Altho our dynamical scheme is the same as the one in \cite{lu}, situation has
changed during the past twenty years. The important improvements happened in the areas
of QCD corrections to the effective weak Hamiltonian 
and in the further development of the Skyrme model.
The latter becomes a good candidate for the correct description of the higher SU(3) representation
multiplets \cite{DPT,SKYRME} and a good tool for evaluation of the nonperturbative quantities
like the 4-quark operator matrix elements between different baryon states.
Comparing the result (\ref{27}) with Table 1 from \cite{lu}, we conclude that the
dynamical scheme is working well in both cases. 
In this work we are using well accepted result for the NLO computations of the QCD corrections  
\cite{bur}. According to \cite{bur} the errors generated 
by different choice of renormalization scheme and scale are less than 10\%.
Overall QCD corrections are more stabile
and smaller today; penguins are about one order of magnitude smaller and also 
$c_1\simeq -3\;\rightarrow -1.9$ \cite{bur}. Skyrme model provide us with 4-quark operator matrix
elements, in different way than the MIT bag model did in \cite{ga,lu}, producing 
rough agreement with measurement (\ref{28}). Finally, according to (e), as a bonus we 
obtain better agreement with experiment for the nonleptonic $\Omega^{-}$ 
and the p-wave hyperon decays.  

Obviously, not all details are under full control. For example $m_{s}$ corrections are
neglected and the uncertainties contained in the input parameters (\ref{23}) are unfortunately large. 
Nevertheless, the QCD-corrected weak Hamiltonian 
$H_{w}^{eff}$, supplemented by the Skyrme model, for the mean values of the input parameters, 
lead to the correct description of the rare $\Omega ^{-} \to \Xi^{0}(1530) \pi^-$ decay. 


\vspace{.1cm}
This work was supported by the Ministry of Science of the
Republic of Croatia under the contract No. 00980102.


\begin{thebibliography}{99}
\bibitem{foot1} In the further we should note $\Xi^{0}_{3/2^+}(1530)\equiv \Xi^*$.
\end{thebibliography}
\end{document}